\documentclass[aps,prb,twocolumn]{revtex4}
\usepackage{amsmath}
\usepackage{graphicx}
\begin{document}

\renewcommand{\thefootnote}{\fnsymbol{footnote}} 

\title{Interaction between Injection Points during Hydraulic Fracturing}

\author{Kjetil M. D. Hals$^{1,\ast}$ and Inga Berre$^{1,2}$}
\affiliation{ $^1$ Christian Michelsen Research, P.O. Box 6031,  NO-5892 Bergen, Norway. \\
 $^2$ Department of Mathematics, University of Bergen, P.O. Box 7800, NO-5020 Bergen, Norway.}

\begin{abstract}
We present a model of the hydraulic fracturing of heterogeneous poroelastic media. The formalism is an effective continuum model that captures the coupled dynamics of the fluid pressure and the fractured rock matrix and models both the tensile and shear failure of the rock.  As an application of the formalism, we study the geomechanical stress interaction between two injection points during hydraulic fracturing (hydrofracking) and how this interaction influences the fracturing process. For injection points that are separated by less than a critical correlation length, we find that the fracturing process around each point is strongly correlated with the position of the neighboring point. 
The magnitude of the correlation length depends on the degree of heterogeneity of the rock and is on the order of $30-45$ m for rocks with low permeabilities. In the strongly correlated regime,  we predict a novel effective fracture-force that attracts the fractures  toward the neighboring injection point.
\end{abstract}

\maketitle

\section{Introduction}
\begin{figure}[ht] 
\centering 
\includegraphics[scale=0.95]{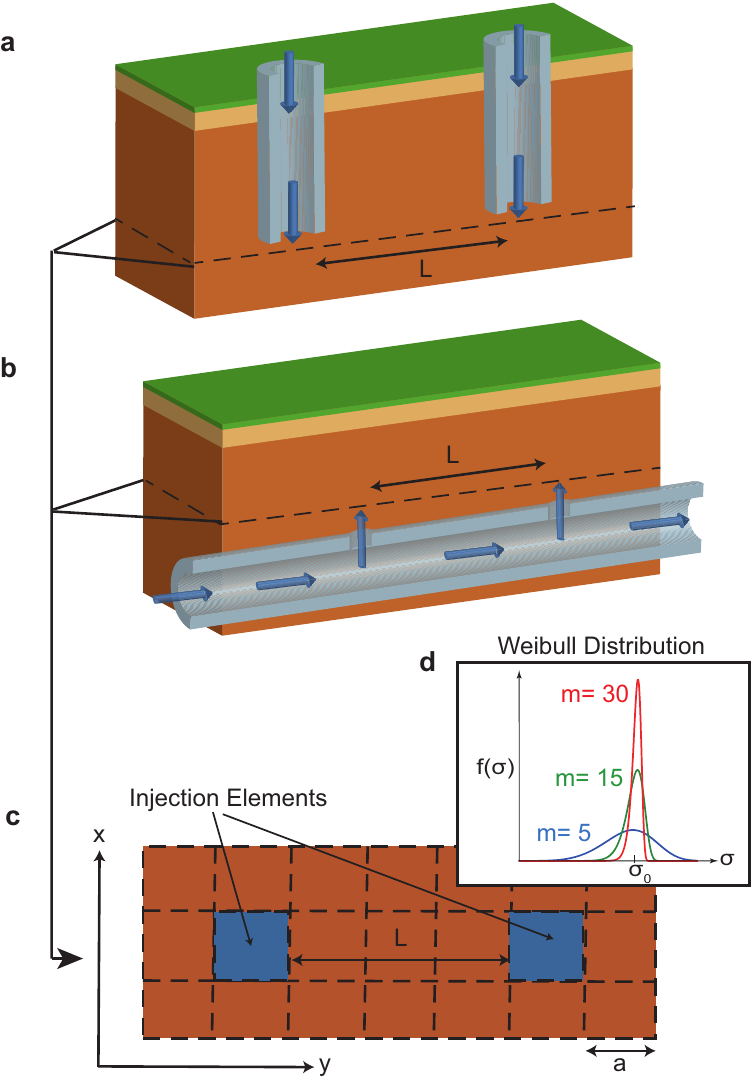} 
\caption{({\bf a}) Hydrofracking using two vertical wells. ({\bf b}) Hydrofracking using a perforated horizontal well. ({\bf c}) A simplified, discretized 2D representation of the reservoirs shown in ({\bf a}) and ({\bf b}). ({\bf d}) A heterogeneous  system is modeled by distributing the material strengths of the discrete volume elements according to the Weibull distribution.  The degree of disorder is tuned by varying the Weibull modulus, m.  }
\label{Fig1} 
\end{figure}

\footnotetext[1]{E-mail: kjetil.hals@cmr.no}
Hydrofracking is a technology that utilizes highly pressurized fluid to create fracture networks in rock layers with low permeabilities. 
A fracking fluid is injected into a cased wellbore, and the parts of the reservoir to be fractured are accessed by perforating the casing at the correct locations (Fig.~\ref{Fig1}b). The injection well can be vertical or horizontal, and several injection points may be active during the hydraulic stimulation of the system (Fig.~\ref{Fig1}a-b). The highly pressurized fluid that flows into the reservoir increases the geomechanical stress around the injection points and causes the rock to fracture. Ideally,  the hydrofracking creates long, distributed cracks that connect a large area of the reservoir to the well. A comprehensive understanding of this fracturing process is therefore crucial to optimize the functionality of the reservoir.  

The creation of a fracture changes the geomechanical strain energy of the system. If the reservoir contains several fractures, the modification of the strain energy mediates an effective interaction between the cracks. The interaction of fractures has been studied theoretically in several works using both direct numerical simulations~\citep{akulich, aghighi} and statistical methods~\citep{masihi,shekhar}. A geomechanical stress interaction also exists between two injection points during hydrofracking. A thorough investigation of this interaction and its consequences for the fracturing process is lacking. 

The failure of homogeneous materials is well understood. A fracture is created when the strain energy release rate exceeds a critical value, and it propagates along the direction that maximizes the energy release~\citep{knott}. The failure of heterogeneous materials, in contrast, depends on many microscopic mechanisms and is challenging to model~\citep{ahansen}. To understand the fracturing process of a heterogeneous material on the micro-scale, statistical models such as the fuse - and the beam- models have been particularly useful~\citep{ahansen,hjherrmann}. In these models, the heterogeneity of the system is taken into account by defining a local, position-dependent material strength that is drawn from a statistical distribution. The fracturing process is therefore not entirely controlled by the energy release rate, but also depends on the disorder of the system and the distribution of the local strengths. Depending on the degree of disorder, the statistical models have shown that heterogeneous systems exhibit fracturing regimes that are distinct from the single-crack development observed in homogeneous systems~\citep{ahansen}.

Several models for hydrofracking have been developed~\citep{jadachi}, but a weakness of most of these models is that they do not account for the heterogeneities of the rock. However, some research based on extensions of the statistical beam model has been applied to the study of hydrofracking of heterogeneous systems~\citep{Tzschichholtz94,Tzschichholtz95,Tzschichholtz98,eflekkoy}. Recently, these studies have been further developed to provide a better description of the rock matrix. In \citet{mwangen} and \citet{swang}, the dynamics of the rock matrix are modeled using poroelastic theory, and the heterogeneity is treated by distributing the local material strength according to a probability distribution.      

In the present paper, we develop a model of the hydrofracking of heterogeneous, poroelastic media. The formalism captures the coupled dynamics of the fractured rock matrix and the fluid pressure. The main parts of the model, which describes the fracturing event and the effective coupling between the pressure and the fractures,  are based on a previous study by \citet{mwangen}.  We extend this work by taking into account the anisotropic fluid flow and the shear failure of the material. Heterogeneities are treated statistically by distributing the local strength of the material according to the Weibull distribution (Fig.~\ref{Fig1}d). We apply the formalism to the study of the mechanical stress interaction between two injection points during hydrofracking, and investigate the dependency of the interaction on the disorder of the rock. For two points that are separated by a distance that is smaller than a critical correlation length, we find that the fracturing process around each injection point is strongly correlated with the position of the neighboring point. The critical correlation length at which this strongly correlated regime occurs depends on the degree of heterogeneity, with correlation lengths of approximately $20$~m for highly disordered systems and $45$~m for  weakly disordered systems. In the correlated regime, we predict a novel {\it effective fracture force} that attracts the fracture toward the neighboring injection point. Our results are important for optimizing the hydraulic stimulation of reservoirs. For well perforations that are separated by a distance that is less than the critical correlation length, the results imply a reduced effect of the stimulation because the fractures are attracted toward neighboring injection points. Knowing the correlation length of the system is therefore crucial for creating an effective and long-ranging fracture network.

This paper is organized in the following manner. In Sec.~\ref{Sec:Theory}, we present the theory and governing equations of our hydrofracking model. The section concludes with a pseudo code of the algorithm. Sec.~\ref{Sec:NumSolAppr} describes the numerical solution strategy. The next two sections apply the formalism to the study of the interaction of two injection points during hydrofracking. Sec.~\ref{Sec:Model} provides a description of the model system that we consider, and Sec.~\ref{Sec:Results} presents our findings. We conclude and summarize our results in Sec.~\ref{Sec:Summary}. 

\section{Governing Equations}
\label{Sec:Theory}
In this section, we consider a poroelastic system and develop the mathematical description that captures the coupled dynamics of the fractured rock matrix and the fluid. 
At the end of the section, the tensile and shear failure criteria are defined. 
 
\subsection{Poroelastic Theory}
Under the action of applied forces, an elastic medium exhibits deformations. The deformations can be in the form of a change in the shape of the object (without a change in its volume), referred to as a shear deformation, or a compression or stretching,  referred to as a volumetric deformation. Let $\mathbf{r}$ denote the position of a material point in the medium before the deformation, and let $\mathbf{r}'$ denote its value after the deformation. The displacement of the point is then given by the  displacement field  $\mathbf{u}(\mathbf{r})= \mathbf{r}' - \mathbf{r}$. The displacement field is a function of the position $\mathbf{r}$, i.e., it describes how each material point moves under a deformation. For small displacements,
the state of the elastic system is completely described by the strain tensor~\citep{landau}:
\begin{eqnarray}
\epsilon_{ij} (\mathbf{r}) &=& \frac{1}{2}\left(  \frac{\partial u_i }{\partial x_j} + \frac{\partial u_j  }{\partial x_i} \right) .
\end{eqnarray}
The diagonal elements of the strain tensor represent local volumetric deformations around each point $\mathbf{r}$, while the off-diagonal terms capture the shear deformations.
In mathematical terms, the displacement field is a vector field, and the strain tensor is a tensor field over the space formed by the elastic object.  

The dynamics of the elastic medium are described by the following fundamental law of elasticity theory~\citep{landau}:
\begin{eqnarray}
\rho \frac{\partial^2 u_i  }{\partial t^2}  &=&  F_i + \frac{\partial \sigma_{ik} }{\partial x_k}. \label{Eq:Biot}
\end{eqnarray} 
Here, $\rho$ is the mass density, $F_i$ represents the external forces, and $\sigma_{ij}$ is the stress tensor arising from the internal stresses. The internal stresses are caused by the intermolecular forces that occur because of the relative displacement of the molecules under the deformation. In Eq.~\eqref{Eq:Biot} (and in what follows), we apply the Einstein summation convention and sum over repeated indices. Usually, the only external force that appears is the gravitational force, $\mathbf{F}= \rho \mathbf{g}$, where $\mathbf{g}$ is the gravitational acceleration. The equilibrium state of the system is determined by solving the stationary equation, i.e., Eq.~\eqref{Eq:Biot} with $\partial^2 u_i / \partial t^2 = 0$. The elastic energy stored in a strained system is~\citep{landau}:
\begin{eqnarray}
E &=& \int  \sigma_{ij}\epsilon_{ij} \text{d}\mathbf{r}
\end{eqnarray}

In this study, we concentrate on a porous, elastic rock system with water-filled pores. In the linear response regime for an isotropic, poroelastic system, the stress tensor can be written phenomenologically as~\citep{jbundschuh}:
\begin{eqnarray}
\sigma_{ij} &=& \sigma_{ij}' - bp_f \delta_{ij}, \label{Eq:stress}
\end{eqnarray}
where
\begin{eqnarray}
\sigma_{ij}' &=& \lambda \epsilon_B \delta_{ij} + 2G\epsilon_{ij}.\label{EffStress}
\end{eqnarray}   
The tensor $\sigma_{ij}'$ is the Terzaghi effective stress tensor, which describes the stresses that act only on the rock matrix, while $\sigma_{ij}$ represents the stresses acting on the total  fluid-rock system;  $\epsilon_B$ is the trace of the strain tensor, and
 $\lambda > 0$ and $G > 0$ are elasticity coefficients, referred to as  the  Lam\'e and rigidity moduli, respectively.  The parameter $b\in [0,1]$ is the effective stress coefficient (also known as the Biot-Willis parameter), which describes how the lithostatic pressure changes with the fluid pressure $p_f$. A pressure $p_f > 0$ corresponds to a fluid pressure that is larger than the atmospheric pressure. We use the following sign conventions: $\epsilon_B > 0$ and $\sigma_{ii} > 0 $ in tension and  $ \epsilon_B < 0$ and $\sigma_{ii} <0$ in compression. Substituting the stress tensor in Eq.~\eqref{Eq:stress} into Eq.~\eqref{Eq:Biot} and expressing the strain tensor in terms of the displacement field produces the equation  of the poroelastic system.
In this paper, we assume that the relaxation time of the elastic system is small compared with the  time scale of the pressure evolution. We can therefore assume that for a given fluid-pressure profile, the elastic medium is always very close to the equilibrium state. Thus, in what  follows, we will be concerned only with the stationary version of Eq.\eqref{Eq:Biot}. 

The above formalism models an isotropic system that does not contain any fractures. In the numerical implementation of the problem, a simple way to model a fracture is to set the elasticity parameters equal to zero (i.e., $G=\lambda=0$) for the discrete volume elements containing a fracture (Fig.~\ref{Fig:FracVol}).  As shown by \citet{mwangen}, this method of modeling the fracture correctly produces the form of the stress field close to the fracture tip. A more correct description of 
the stress field close to the fracture tip requires a time-dependent grid with an extra fine mesh size near the fracture tip. Such a detailed description is beyond the scope of the present model, in which the aim is
to provide a qualitative description of the fracturing process.  

\subsection{Pressure Equation} 
\label{Sec:Pressure}
The equation for the fluid-pressure is derived from the following continuity equation based on fluid-mass conservation~\citep{jbundschuh}:
\begin{eqnarray}
\frac{\partial \zeta}{\partial t}   &=& - \boldsymbol{\nabla}\cdot \mathbf{v}_f + Q (\mathbf{r},t)  . \label{pressureEq1}
\end{eqnarray} 
Here, $\zeta(\mathbf{r}, t)$ is the fluid content added
to the bulk volume, and $Q (\mathbf{r},t)$ is a source (sink) term that arises from the injection (extraction) of the fluid. The quantity $\mathbf{v}_f$ is the Darcy velocity, which can be expressed in terms of the fluid pressure ($p_f$), the permeability ($\mathbf{k}$), the viscosity ($\mu_f$), and the gravitational force ($\rho_f\mathbf{g}$, where $\rho_f$ is the mass density of the fluid): $\mathbf{v}_f = -(\mathbf{k}/\mu_f) (\boldsymbol{\nabla} p_f - \rho_f \mathbf{g})$.  In the most general (anisotropic) case, the permeability $\mathbf{k}= [k_{ij}]$ is a tensor. In particular, the permeability is strongly anisotropic in the parts of the solid that contain fractures.  

Let us first consider a homogeneous poroelastic medium that does not contain any fracture zones. In this case, there are two independent mechanisms that contribute to the fluid content $\zeta$: an increase in the fluid pressure without an increase in the bulk strain (i.e., the fluid molecules are more densely packed), and an increase in the 
pore volume caused by an increase in the bulk volume (while $p_f$ remains constant). This implies that  the fluid content can be written as~\citep{jbundschuh}:
\begin{eqnarray}
\zeta &=& S p_f  + b \epsilon_B, \label{zeta}
\end{eqnarray}
where $\epsilon_B$ is the volumetric change of the bulk volume. The parameter $S$ is the constrained specific storage factor. 
Thus, the time variation of $\zeta$ has two contributions: one term that is proportional to $\partial p_f / \partial t$ and a second term that is proportional to $\partial \epsilon_B  / \partial t$.
The characteristic time scale ($t_{\epsilon}$) of the last term strongly depends on the rock type, and the time scale of the first term ($t_p$) is primarly governed by the compressibility of the fluid.
To compare these two time scales, let us consider an elastic medium that is not confined by any external forces so that it is free to expand/contract in response to a pressure perturbation.
The response of the bulk volume is then given by $\delta \epsilon_B = H^{-1}\delta p_f$, where $H^{-1}$ is the special poroelastic expansion coefficient ~\citep{jbundschuh}. With this expression for the volumetric response, we obtain the following ratio between the two time scales:
\begin{eqnarray}
\frac{t_{\epsilon}}{t_p} &=& \frac{b}{S H}.\label{Eq:TimeScale}
\end{eqnarray}
For hard rocks with a low porosity, i.e.,  $\phi < 0.1$ and $b  <  0.4$, this time ratio is small, $t_{\epsilon}/t_p < 0.1$~\citep{jbundschuh}. Thus, for hard-rock systems, the second term in Eq.~\eqref{zeta} is negligible compared with the first term, and we therefore disregard it in the following equations. This corresponds to assuming that the porous medium is incompressible. We also assume that the fluid is slightly compressible. In this approximation, the constrained storage factor is $S= \phi \beta_f$, where $\beta_f = \rho_f^{-1} (\partial \rho_f / \partial p_f)$ is the compressibility factor under isothermal conditions.  Eq.~\eqref{pressureEq1} then becomes
\begin{eqnarray}
\phi \beta_f \frac{\partial p_f}{\partial t}   & = &  \boldsymbol{\nabla} \cdot \frac{\mathbf{k}  }{\mu_f  } (\boldsymbol{\nabla} p_f - \rho_f \mathbf{g}) + Q(\mathbf{r},t)  . \label{pressureEq2}
\end{eqnarray}

\begin{figure}[ht] 
\centering 
\includegraphics[scale=1.0]{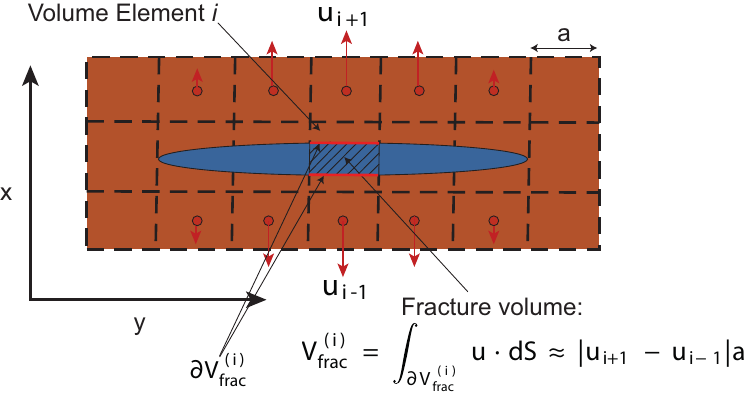} 
\caption{The figure shows a system containing a fracture (the blue area). The fracture volume is calculated numerically from the displacement field in the neighboring volume elements of the fracture. The displacement field is illustrated by the red arrows.  The elasticity parameters of the elements containing parts of the fracture are equal to zero. The volume of each element is
$V_0^{(i)}= a^2$.  }
\label{Fig:FracVol} 
\end{figure}

Next, we treat the equation describing the pressure of  the fracture zones. 
In zones with fractures, a pressure gradient opens the fractures and consequently modifies the local permeability and porosity. 
The time variation of this process is of the same order as the pressure variation, and is a non-negligible contribution to the pressure evolution. As a consequence, the pressure equation
contains terms that couple the pressure dynamics to the fracture dynamics.  In addition, the permeability becomes anisotropic with a much higher permeability along the fractures than across the fractures.
As described by \citet{mwangen}, one method to include this coupling in a numerical implementation of the problem is to assign to each fractured volume element (of the discretized system) an effective porosity that depends on the opening of the fracture (Fig.~\ref{Fig:FracVol}). 
The effective porosity of the volume element $i$ is:
\begin{eqnarray}
\phi^{(i)}_{\text{eff}} (t) &=& \phi + (1-\phi)\phi_{\text{frac}}^{(i)} (t),
\end{eqnarray}
where 
\begin{eqnarray}
\phi_{\text{frac}}^{(i)} (t) &=& V^{(i)}_{\text{frac}} (t)/V^{(i)}_0.
\end{eqnarray}
Here, $V^{(i)}_{\text{frac}} (t)$ is the volume of the fracture inside element $i$, and  $V^{(i)}_0$ is the volume of element $i$. 
The fracture volume of each element is found by integrating the displacement field over the fracture surface (Fig.~\ref{Fig:FracVol}): 
\begin{eqnarray}
 V^{(i)}_{\text{frac}} (t)&=& \int_{\partial V^{(i)}_{\text{frac}}} \mathbf{u} (t) \cdot \rm d\mathbf{S}. \nonumber
 \end{eqnarray}
The effective porosity
becomes equal to one if the fracture fills the entire volume element, and it reduces to the porosity of the rock if the fracture is closed or if element $i$ does not contain a fracture.

Incorporating the effects of the fractures with an effective porosity is a suitable approximation as
long as the typical length scale of the pressure variations is large compared with the opening of the fracture. 
Let $\phi_{\text{eff}}(\mathbf{r},t)$ denote the continuum limit of the effective porosity. 
Because of the time variation of the effective porosity, the continuity equation for the fluid mass now becomes: 
\begin{eqnarray}
\phi_{\text{eff}} \frac{1}{\rho_f}\frac{\partial \rho_f}{\partial t} + \frac{\partial \phi_{\text{eff}}}{\partial t} &=& - \boldsymbol{\nabla}\cdot \mathbf{v}_f + Q, 
\end{eqnarray}
which yields the following the pressure equation:
\begin{eqnarray}
\phi_{\text{eff}} \beta_f \frac{\partial p_f}{\partial t}  +  \frac{\partial \phi_{\text{eff}} }{\partial t } & = & \boldsymbol{\nabla} \cdot \frac{\mathbf{k}  }{\mu_f  } (\boldsymbol{\nabla} p_f - \rho_f \mathbf{g}) + Q. \label{Eq:pressureEq3}
\end{eqnarray}

The fractures also modify the permeability, which becomes an anisotropic second-rank tensor. In our numerical implementation of the problem, we assign to each volume element a permeability tensor that depends on the  fracture direction and the fracture opening (In the numerical model, we allow the fracture to propagate only in the x and the y directions for simplicity).    As an illustration, let us consider the situation shown in Fig.~\ref{Fig:FracVol}. This 2D system has an open fracture along the y-axis, and the permeability is therefore larger along the y-direction than along the x-axis. We model this effect by introducing the following tensor:
\begin{eqnarray}
\mathbf{k}^{(i)} &=&  
\begin{pmatrix}
k_{x}^{(i)} & 0 \\
0 & k_{y}^{(i)}
\end{pmatrix}
=
\begin{pmatrix}
k_{\bot}^{(i)} & 0 \\
0 & k_{||}^{(i)}
\end{pmatrix}, \label{Eq:perm}
\end{eqnarray}
where $k_{\bot}^{(i)}\approx k_{\text{rock} }$ is the permeability across the fracture, which is equal to the permeability of the rock, and $k_{||}^{(i)}\approx k_{\text{rock}} + (k_{\text{fluid}} - k_{\text{rock} } )\phi_{\text{frac} }^{(i)}$
is the permeability along the fracture. The quantity $k_{\text{fluid}} $  represents the permeability inside the fracture.  In the parallel plate model for a single fracture, the fracture permeability is given by the cubic law $k_{\text{fluid}}= w^2/12$ where $w$ is the fracture aperture. This yields a very large permeability that may cause numerical problems. As mentioned by \citet{mwangen}, a practical solution to this problem is to choose a permeability that is large enough to enact a minimal pressure drop along the fracture, but is small enough to avoid numerical instabilities. 
Eq.~\eqref{Eq:perm} becomes $\mathbf{k}= k_{\text{rock}}\mathbf{I}$ ($\mathbf{I}$ is the identity matrix) for elements that are not fractured or that contain a closed fracture.
The permeability tensor, as defined in Eq.~\eqref{Eq:perm}, is position-dependent and therefore gives rise to terms that are proportional to $\boldsymbol{\nabla} k_{ij}$ in Eq.~\eqref{pressureEq2}.

\subsection{Fracture Criteria}
Porous materials fail under the action of a large fluid-pressure gradient, and   
the failure can be induced by both shear and tensile forces. 
Several phenomenological failure criteria exist~\citep{jcjaeger,mspaterson,hjherrmann}. To capture both tensile and shear failure, we adopt two distinct criteria.  

We model the tensile failure by locally defining (i.e., for each volume element $i$) a critical tensile stress, $\sigma_c^{(i)} > 0$. If one of the eigenvalues  of the Terzaghi effective stress tensor (evaluated at element $i$)  exceeds the critical stress, the volume element $i$ fails. The compressive strength of the material can be included in a similar manner. However, during hydraulic fracturing,  the tensile failure is the dominant fracturing mechanism, and we therefore disregard compressive failure in our numerical simulation.

The shear failure is modeled in 2D by the Mohr-Coulomb criterion~\citep{jcjaeger,mspaterson}:
\begin{eqnarray}
 |  (\sigma_1^{(i)} - \sigma_2^{(i)} ) / 2  | &\geq& -\frac{ \sigma_1^{(i)} + \sigma_2^{(i)} }{2} \sin\theta  +  c^{(i)}\cos \theta . \nonumber
\end{eqnarray}
Here, $\theta$ is the internal friction angle, which depends on the density, surface, and shape of the particles constituting the
solid, and the cohesion, $c$, describes the minimal shear force that is required for fracturing when no normal stresses are present.  
Furthermore, $\sigma_{\alpha}^{(i)}$ are the eigenvalues of the Terzaghi effective stress tensor evaluated for volume element $i$. 
In a 3D system, the Drucker-Prager criterion is used~\citep{jcjaeger}:
\begin{eqnarray}
\tau^{(i)}  & \geq & \frac{2}{ (1 + \alpha)^2 }  \left( -\left( \sigma_1^{(i)} + \sigma_2^{(i)} + \sigma_3^{(i)} \right) +  \alpha \sigma_c^{(i)} \right)^2, \nonumber
\end{eqnarray}       
where $\tau^{(i)}$ is:
\begin{eqnarray}
\tau^{(i)}  &\equiv&   (\sigma_1^{(i)} - \sigma_2^{(i)})^2  + (\sigma_3^{(i)} - \sigma_2^{(i)})^2   +  (\sigma_1^{(i)} - \sigma_3^{(i)})^2 . \nonumber
\end{eqnarray}
The parameter $\alpha > 1$ is a dimensionless material parameter. 

As discussed by \citet{mwangen}, the critical stresses (i.e., the parameters $ \sigma_c^{(i)}$ and $c^{(i)}$ in our model) depend on the grid size because the stress field is singular at the fracture tip.
\citet{mwangen} solves this problem by scaling the critical stress with the square root of the grid size. An alternative procedure, is to use experimental data to fit the values for 
$\sigma_c^{(i)}$ and $c^{(i)}$, such that fracturing occurs for fluid pressures in the experimental range. In this paper, we apply the latter approach by tuning the mean values of
$\sigma_c^{(i)}$ and $c^{(i)}$ so that shear and tensile failure appear for bore-hole pressures  that are typical for the present problem.

The strength of a material usually follows a distribution that is well-described by the Weibull distribution (see Fig.\ref{Fig1}d)\citep{hjherrmann}:
\begin{eqnarray}
f(\sigma)  &=&  \frac{m}{\sigma_0} \left(  \frac{\sigma - \sigma_{th}}{\sigma_0} \right)^{m-1}  \exp \left( - \left(  \frac{\sigma - \sigma_{th}}{\sigma_0} \right)^m   \right) . \nonumber
\end{eqnarray} 
Here, $m$ is the Weibull modulus, $\sigma_0$ is the mean value of the critical stress, and $\sigma_{th}$ is the threshold stress below which no failure will occur (usually, it is set as $\sigma_{th}=0$). 
We incorporate the heterogeneity of the rock into our model by distributing the local strengths $\sigma_c^{(i)} > 0$ and $c^{(i)} > 0$ according to the Weibull distribution. 
The Weibull modulus is a measure of the degree of disorder in the system. A system with  a small Weibull modulus has a higher degree of disorder than a system with a larger modulus.
This method of modeling the heterogeneity is analogous to what is done in microscopic fracturing models, such as the fuse - and the beam -models.

\section{Numerical Solution Approach}
\label{Sec:NumSolAppr}
The model presented in the previous section is an effective continuum model of a fractured poroelastic system.
Our primary aim is not to provide a detailed description of the stress field close the fracture tip or the fluid flow inside the fractures. Instead, we seek a simplified description of the coupled fluid-rock system that captures the primary elements of the fracturing process to obtain a qualitative understanding of hydrofracking, for example, to sort out the mechanisms that dominate the process or to map out what types of fracturing regimes are produced by different material parameters.

To solve the system of equations presented in the previous section, we apply a sequential solution strategy using standard numerical discretization methods.  

\subsection{Discretization Methods}
The pressure equation, Eq.~\eqref{Eq:pressureEq3}, is solved with a finite difference formulation of the complete pressure equation that captures the pressure dynamics in the fracture zones and in the homogeneous regions of the system.  
To isolate the effect of interactions between the two injection points, we disregard the gravitational force in the Darcy velocity, and for simplicity, we consider a 2D rectangular region that is discretized with a lattice constant, $a$, as illustrated in Fig.~\ref{Fig1}c. An explicit scheme is implemented using a forward Euler discretization in time for the temporal pressure derivative. The time derivative, $\partial \phi_{\text{eff}}  / \partial t$, is calculated from the last two time steps.

The ordinary differential equations arising from the discretized pressure equation are solved using the Cash-Karp embedded Runge-Kutta method~\citep{press}, which is an adaptive algorithm that regulates the time step using an error estimate. 

The boundary condition for the pressure is $p_f=0$. The Galerkin method with bilinear trial functions is used to solve the
stationary  stress equation, Eq.\eqref{Eq:Biot} (see \citet{hplangtangen} for more details). We use traction-free boundaries as boundary conditions in the present problem, 
 i.e., $\int_{\partial \Omega} \sigma_{ij}n_j \text{dS}= 0 $ for all $i$, where $\partial \Omega$ is the boundary of the system, and $\mathbf{n}$ is its surface normal.

Our simplified method of modeling the fractures results in a grid-dependent fracture aperture and effective porosity. 
In practice, we fix the grid size by claiming that the opening of the fracture is of the order $1$ mm when the pressure inside the fracture is of the order $1$ MPa.

\subsection{The Fracture Event}
\label{sec:FracEvent}
The present model assumes that the reservoir consists of hard rocks with a low porosity, so that the ratio in Eq.~\eqref{Eq:TimeScale}  is small.
Physically, this means that the relaxation time of the rock system is much less than the time variations of the pressure distribution.
We can therefore assume that the fracture event happens instantaneously. This leads to a sudden fluid-pressure drop inside the fracture, while the pressure outside the fracture is not affected. The new fluid pressure inside the fracture 
is determined using the Newton iteration, based on the assumption that the mass of fluid inside the fracture is preserved during the fracture event. The iterative scheme is:  \\

\begin{enumerate}
\item Adjust $p_f\mapsto p_f^{\text{new}}$.
\item Solve stationary Eq.~\eqref{Eq:Biot} with the new pressure profile.
\item Calculate the new fracture volume, $V_{\text{frac}}^{\text{new}}$.
\item If $V_{\text{frac}}^{\text{old}}\rho(p_f^{\text{old}},T)=V_{\text{frac}}^{\text{new}}\rho(p_f^{\text{new}},T)$, stop the iteration process. If not, return to step 1. \\
\end{enumerate} 

Here, $\rho(p,T)$ is the equation of state of the fluid, which expresses the fluid-mass density as a function of the temperature and the pressure.  

\subsection{The Algorithm}
We end this theory section with a brief pseudo code of the numerical solution strategy: \\

\begin{enumerate}
\item Solve the pressure equation, Eq.~\eqref{Eq:pressureEq3}, for the current time step. 
\item Solve the stationary  stress equation, Eq.~\eqref{Eq:Biot}, with the new pressure profile.
\item Check the fracture criteria for each volume element.
\item If no fracturing occurs:
\begin{enumerate}
\item Calculate the new fracture volumes. Update $\phi_{\text{eff}}^{(i)}$, $\partial \phi_{\text{eff}}^{(i)} / \partial t$, and $\mathbf{k}^{(i)}$.
\item Calculate a new time step from the error estimate, and return to step 1.
\end{enumerate} 
\item If fracturing occurs:
\begin{enumerate}
\item Set $G=\lambda= 0$ for the fractured volume elements.
\item Find the new fluid pressure and the volume of each fracture using the iterative scheme in Sec. \ref{sec:FracEvent}. Update $\phi_{\text{eff}}^{(i)}$, $\partial \phi_{\text{eff}}^{(i)} / \partial t$, and $\mathbf{k}^{(i)}$.
\item  Calculate a new time step from the error estimate, and return to step 1. Use the updated pressure profile  as the initial pressure for the next time step.
\end{enumerate} 
\end{enumerate}

\section{Model Specifications}\label{Sec:Model}
Next, we apply this fracturing model to the study of the geomechanical stress interaction between two injection points during the hydrofracking of a rock system with low permeability.

The elasticity parameters of the system are $G= 13.7$~GPa and $\lambda= 21.7$~GPa. The mass-density of the fluid-rock system is $\rho = 2620$ $\rm kg/m^3$. The effective stress coefficient is $b= 0.38$, and the fluid compressibility is $\beta_f= 5.9\times 10^{-10}$ $\rm Pa^{-1}$. The viscosity of the fluid, the porosity and the
permeability of the rock are  $\mu_f=200\times10^{-6}$ Pa\;s, $\phi= 0.08$ and $k_{\rm rock}= 0.026$~mD, respectively. The viscosity and the compressibility of the fluid correspond to that of pressurized water (in the liquid phase) at a temperature of approximately $413$ K~\citep{jbundschuh}. These poroelastic parameters are collected from the technical data of the Los Humeros geothermal field~\citep{jbundschuh}.  With these material parameters, the time ratio in Eq.~\eqref{Eq:TimeScale} is $t_{\epsilon}/ t_p \approx 0.1 $. The permeability, $k_{\rm fluid}$, inside the fractures is four orders of magnitude larger than the rock permeability. 

We represent the reservoir as a 2D discretized system. 
To solve the pressure and stress equations, i.e., Eq.~\eqref{Eq:pressureEq3} and Eq.~\eqref{Eq:Biot}, we use a quadratic grid, as illustrated in Fig.~\ref{Fig1}c, with a grid size of $a= 5$ m. 
The dimensions of the system range from $L_x=L_y= 150-220$~m, depending
on the separation of the injection points. 

The simulation starts with two injections points separated by a distance, $L$. Initially, the system contains no fractures. The elasticity parameters of the injection elements are equal to zero. 
The mean values for the cohesion, $c$, and the critical tensile stress, $\sigma_c$, are $1$~MPa, and the internal friction angle is $40$ degrees. These values result in tensile and shear failure for bore-hole pressures, typically of the order of
$1-8$~MPa. 
We consider systems with a Weibull modulus of $m\in \left\{ 5, 15, 30 \right\}$ (Fig.~\ref{Fig1}d). The injection points are placed along the
$y$-axis, as illustrated in Fig.~\ref{Fig1}.

In the present paper, our main aim is to investigate how the stress interaction between two injection points in combination with disorder influences the hydrofracking process.
To isolate this effect, we have disregarded the gravitational force in our numerical implementation. However, in the next section, we provide a discussion of the
consequences of the gravitational force and its interplay with the stress interaction.

\section{Results and Discussion}\label{Sec:Results}
\begin{figure}[ht] 
\centering 
\includegraphics[scale=1.0]{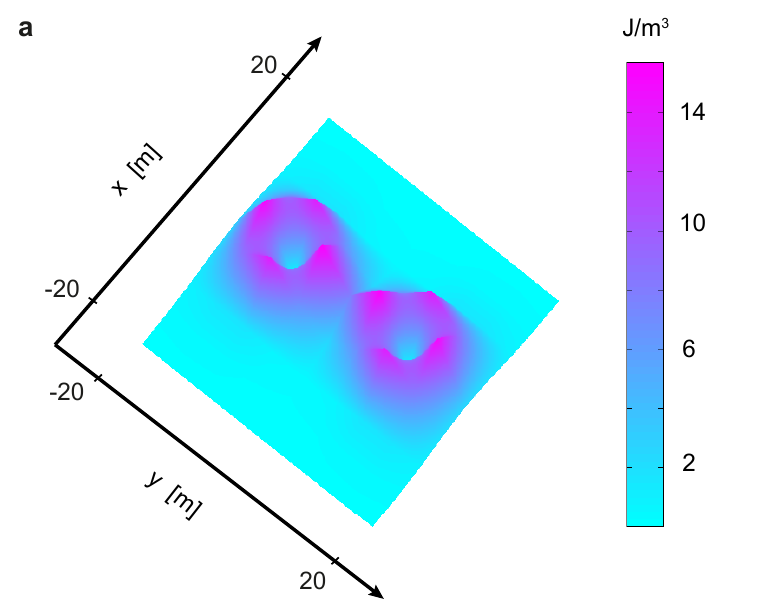}
\includegraphics[scale=1.0]{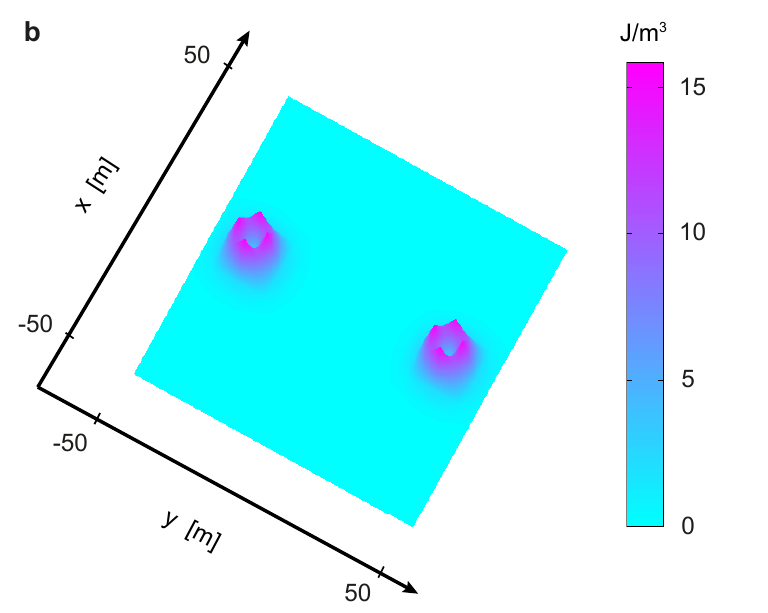} 
\caption{
({\bf a}) Elastic energy density of a system containing two injection points separated by a distance of $15$~m. The bore-hole pressure is $5$~MPa, and the injection rate is $10^{-4}$~$\rm m^3/s$. 
({\bf b}) Elastic energy density of a system containing two injection points separated by a distance of $65$~m. The bore-hole pressure is $5$~MPa, and the injection rate is $10^{-4}$~$\rm m^3/s$.
Note that the rotational symmetry of the system is broken because the injection element is quadratic, i.e. the rotational symmetry around the injection point is reduced to quadratic symmetry.
This is why the strain energy in ({\bf b}) is not isotropically distributed around each injection point. }
\label{Fig2} 
\end{figure}
Eq.~\eqref{EffStress} shows that geomechanical stresses arise from spatial variations of the displacement field.
Large spatial variations result in large strain and stress fields. During hydrofracking, the stress field is largest close to the injection point and relaxes towards 
zero farther away from the point (in the absence of gravity).  Far from the injection point, the displacement field is equal to zero. The length scale over which the displacement field relaxes toward a constant vector field is referred to as the relaxation length. If two injection points are separated by a distance less than twice the relaxation length, the displacement field around each point is influenced by the presence of the neighboring point.  This leads to an effective stress interaction between the two points. 

A consequence of the geomechanical stress interaction is that the strain energy increases in the area between the two points.  Fig.~\ref{Fig2}a-b shows the elastic energy density, $\sigma_{ij}\epsilon_{ij}$, stored in the elements close to the two injection points when the bore-hole pressure is $5$~MPa. In Fig.~\ref{Fig2}a, the injection points are separated by $15$~m. In this case, there is a significant stress interaction between the two points, which causes the elastic energy density to be largest in the region  between the  points.  The rock here is under particularly strong tensile stress. In Fig.~\ref{Fig2}b, the points are separated by $65$~m. For such a large separation, 
the interaction between the points becomes negligible, and the elastic energy density is equally distributed around each of the two injection points. 

In homogeneous materials under tensile stress, a fracture propagates along the direction that causes the largest strain energy release rate. During hydrofracking, we therefore expect the stress interaction between two injection points to mediate an effective force on the fractures that are created close to one of the points. This effective fracture force is expected to drive the fractures toward the neighboring point because this leads to the highest release rate of elastic energy. In heterogeneous materials, this effective force is accompanied by disorder effects. 
Whether the fracturing process is disorder-driven or effective force-driven, i.e., as the dominant fracturing mechanism, depends on the degree of disorder.  

Let us, at this point, briefly discuss the effect of the gravitational force. The gravitational force yields a large compressive stress of the order $-25$ MPa per distance of $1000$ m beneath the surface, which is approximately one order of magnitude 
larger than the tensile stresses produced by the injection of fluid. The effect of the gravitational force is to align the fractures along $\mathbf{g}$. 
Thus, there are two distinct forces acting on the fractures: 
one caused by the gravitational field that drives the fractures in the vertical direction, and one arising from the geomechanical stress interaction that drives the fractures toward neighboring injection points.
The effect of the gravitational field is well-known. In contrast, the effect of the geomechanical stress interaction is new and is the focus of the present paper.

To investigate how the stress interaction influences the fracturing process of heterogeneous systems, we calculate the ensemble-averaged propagation direction, $\langle \mathbf{n} \rangle$.
The vector, $\mathbf{n}= (n_x\  n_y)$, is a unit vector that denotes the initial propagation direction for a fracture created from one of the two injection points. By definition, $\mathbf{n}= (0\ 1)$ points toward 
the neighboring point.  The value of $\mathbf{n}$  is a function of the micro-state of the system. In other words, its value depends on the distribution of the local material strengths and the heterogeneity of the system.  
To map the disorder dependency of the fracturing process, we ensemble-average, $ \mathbf{n} $, by averaging over several micro-states. 
A non-vanishing $\langle \mathbf{n} \rangle$ implies that a fracture has a larger probability of propagating along this direction than in the other directions. In this case, the fracturing process is strongly influenced by the stress interaction and the location of the neighboring injection point. In contrast, a vanishing $\langle \mathbf{n} \rangle$ implies an isotropic distribution of $\mathbf{n}$, and the governing fracturing mechanism is the disorder of the system.  
\begin{figure}[ht] 
\centering 
\includegraphics[scale=1.0]{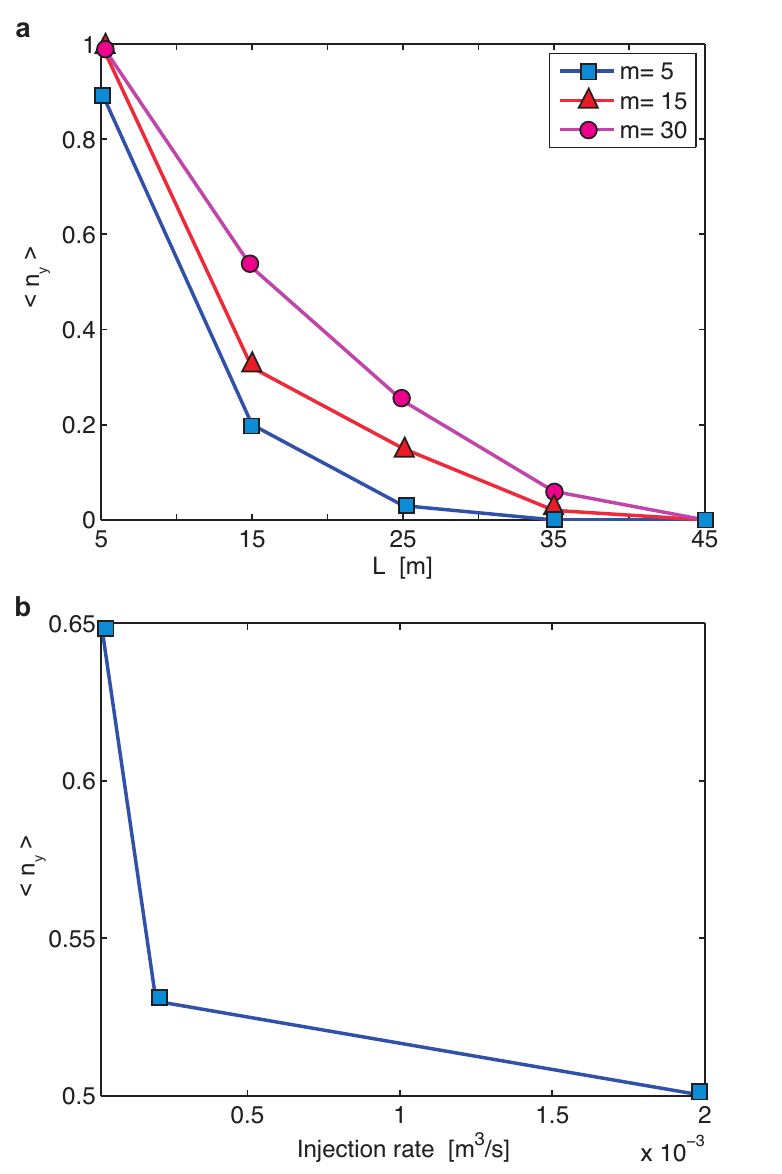} 
\caption{ ({\bf a}) The ensemble average, $\langle n_y \rangle$,  as a function of the separation of two injection points. $\langle n_y \rangle$ is studied for different Weibull moduli, $m$. The injection rate is  $2\times 10^{-4}$~$\rm m^3/s$.  ({\bf b})  The ensemble average, $\langle n_y \rangle$, for the injection rates $\left\{2\cdot 10^{-5},\  2\cdot 10^{-4},\  2\cdot 10^{-3}  \right\}$~$\rm m^3/s$. The separation between the two injection points is $15$~m, and the Weibull modulus is 30. In both plots, the lines are guides to the eye, and the ensemble average is obtained by averaging over 100 statistical realizations of the system.  } 
\label{Fig3} 
\end{figure}
Fig.~\ref{Fig3}a shows  $\langle n_y \rangle$ as a function of the separation length, $L$, of the two injection points. The injection points are placed along the y-axis. $\langle n_x \rangle$ is not shown because it is equal to zero for the systems we consider.  As expected,  $\langle n_y \rangle$ decreases as the separation becomes larger.   This is because of a weaker interaction between the two injection points, which results in a weaker effective fracture force and a reduced probability for the fracture to propagate toward the neighboring point. How quickly $\langle n_y \rangle$ decays depends primarily on the degree of disorder. For a 
Weibull modulus of $30$, there exists an effective attraction toward the neighboring point for a separation less than $45$ m. We refer to this separation as the critical correlation length. If the Weibull modulus is  $5$, which corresponds to a highly disordered system, the critical correlation length decreases to approximately $30$ m. For separations less than the critical correlation length, the fracturing process around each injection point is strongly correlated with the position of the neighboring point, and the fracturing process is governed by the effective fracture force. For separations larger than the critical correlation length, the disorder effects dominate the fracturing process.

Fig.~\ref{Fig3}b shows  $\langle n_y \rangle$ as a function of the injection rate for a system in which the Weibull modulus is $30$, and the two injection points are separated by $15$ m. 
A decreasing injection rate results in a larger $\langle n_y \rangle$ value. This is  because of a stronger stress interaction between the two points. The stronger interaction arises because of a slower pressure build-up at the injection points, which leads to a longer time interval before the fracturing appears compared with a system with a higher injection rate. The pressure therefore diffuses a larger distance into the rock medium before the fracturing appears, which results in a larger relaxation length of the displacement field.   The larger relaxation length yields a larger stress interaction between the points. 

\begin{figure}[ht] 
\centering 
\includegraphics[scale=1.0]{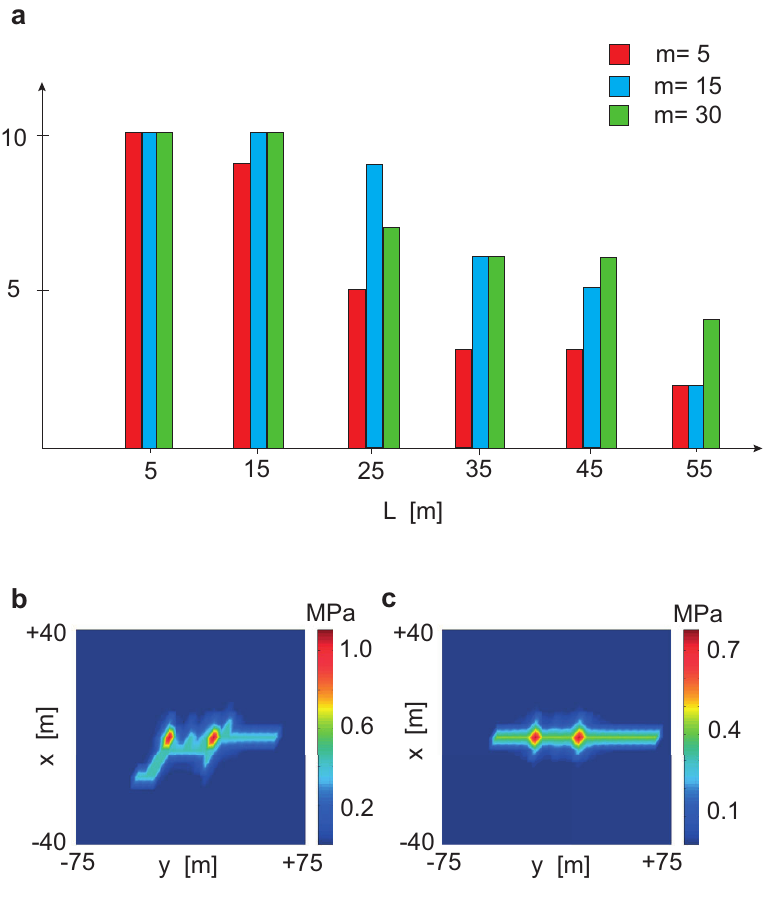} 
\caption{ ({\bf a}) The histogram shows how many times the fracturing process creates a connecting fracture between the two injection elements. For each length, $L$, and Weibull modulus, $m$, 10 fracturing events are simulated.
({\bf b}) The fluid pressure along a connecting fracture between two injection points in a system with a Weibull modulus of $5$.   
({\bf c}) The fluid pressure along a connecting fracture between two injection points in a system with a Weibull modulus of $30$. 
In ({\bf b}) and ({\bf c}), the injection points are separated by $25$ m. }
\label{Fig4} 
\end{figure}
The effective fracture force enhances the chance to create a connecting fracture network between two injection points. 
Fig.~\ref{Fig4}a shows how many times a number of fracturing processes creates a connecting fracture between the two injection points as a function of the separation for systems
with different degrees of disorder. Two examples of a connecting fracture are shown in Fig.~\ref{Fig4}b and Fig.~\ref{Fig4}c for systems
with Weibull moduli of $5$ and $30$,  respectively. In the typical time evolution of such a fracture,  the fracture first connects the two points before
a new fracture arm is created from each point. In addition, for a system with a high Weibull modulus (Fig.~\ref{Fig4}c), the fracture is most likely to show the same character as a fracture in a  homogeneous system, 
i.e.,  a straight line fracture, while the fractures in highly disordered systems  (Fig.~\ref{Fig4}c) propagate more randomly. 
The ability to create a connecting fracture network is crucial for creating a geothermal system, such as an Enhanced Geothermal System (EGS), as well as in  the exploitation of unconventional hydrocarbon resources.  In the case of hydrofracking, the effect of the stress interaction is important. During hydrofracking,  the perforation zones ideally are completely uncorrelated to create long and deep cracks that 
connect a large area of the reservoir to the well. If the zones are spaced by a distance of less than the critical correlation length, the stress interaction leads to a reduced hydraulic stimulation of the system. 
We therefore believe that our results  can be used in the optimization of the hydrofracking process.

\section{Conclusions}\label{Sec:Summary}
In this paper, we developed a model of hydrofracking and applied the formalism to the study of how geomechanical stress interactions between two injection points influence the fracturing process.
We found that when the separations between the two injection points is less than a critical correlation length, the fracturing process around each injection point is strongly correlated with the position of the neighboring point.
The magnitude of the critical correlation length depends on the degree of heterogeneity of the rock. For weakly disordered systems, the correlation length can be as large as $45$~m, and for highly disordered rock systems, it reduces to approximately $20$~m.  In the strongly correlated regime, there exists an effective fracture force that drives the fractures toward the neighboring injection point. An important observation in this work is that the fracture force 
reduces the effectiveness of the hydraulic stimulation if the injection points are separated by a distance less than the critical correlation length.

\section{Acknowledgments}
We are grateful to \O . Pettersen at Uni CIPR for stimulating discussions and comments on the manuscript.

\end{document}